\documentclass[prl,onecolumn,preprintnumbers,nofootinbib,showpacs,amsmath,amssymb,superscriptaddress]{revtex4}

\usepackage[pagebackref=false,
	  colorlinks=true,
      linkcolor=blue,
      urlcolor=blue,
      filecolor=black,
      citecolor=red,
      pdfstartview=FitV,
      pdftitle={},
        pdfauthor={},
        pdfsubject={},
        pdfkeywords={},
        pdfpagemode=None,
        bookmarksopen=true]{hyperref}

\usepackage{amsfonts}
\usepackage{amsmath}
\usepackage{amssymb}
\usepackage{upgreek}
\usepackage{bm}
\usepackage{dcolumn}
\usepackage{epsfig}
\usepackage{graphicx}
\usepackage{graphics}
\usepackage[latin1]{inputenc}
\usepackage{latexsym}
\usepackage{rotating}
\usepackage{tipa}
\usepackage{hyperref}
\usepackage[all]{hypcap}
\usepackage{xspace}

\def \be {\begin{equation}}
\def \ee {\end{equation}}
\def \bea{\begin{eqnarray}}
\def \eea{\end{eqnarray}}
\def \ba {\begin{align}}
\def \ea {\end{align}}

\def \a {\alpha}
\def \b {\beta}

\def \r {\rho}

\def \f {\frac}

\def \nn {\nonumber}

\def \la {\leftarrow}
\def \ra {\rightarrow}

\def \wt {\widetilde}

\def \la {\label}

\begin{document}

\title{Entropic corrections and modified Friedmann equations in the emergence of cosmic space}

\author{Fang-Fang Yuan}
\email{ffyuan@emails.bjut.edu.cn}
\affiliation{Institute of Theoretical Physics, Beijing University of Technology,
Beijing 100124, China}
\author{Yong-Chang Huang}
\email{ychuang@bjut.edu.cn}
\affiliation{Institute of Theoretical Physics, Beijing University of Technology,
Beijing 100124, China}


\begin{abstract}
In the emergent cosmic space scenario recently proposed by Padmanabhan,
the expansion of the universe is attributed to the difference between the
number of degrees of freedom on a holographic surface and the one in the bulk.
Based on this idea, we explore the cases where logarithmic and power-law entropic corrections are present, respectively.
The previously found modified Friedmann equations are reproduced through an improved proposal,
and we show that the same results can be obtained using a generalized expansion law.
\end{abstract}


\maketitle

{\emph{Introduction.}}---
The emergent gravity paradigm \cite{Padmanabhan:2009vy} has been investigated for a long time since the seminal work of Sakharov \cite{Sakharov:1967pk}.
Its reinvigoration in recent years was mainly because of Verlinde's proposal \cite{Verlinde:2010hp}
that gravity can be identified with an entropic force caused by changes in
the information associated with the positions of material bodies.
For related works, see e.g. Refs. \cite{{Jacobson:1995ab},{Padmanabhan:2009kr}}.
Via the relation to thermodynamics, all these studies indicated that
the gravitational field equations can be interpreted as the equations of an emergent phenomenon.

Recently, Padmanabhan \cite{Padmanabhan:2012ik} proposed a novel idea
that the cosmic space itself can also be emergent as cosmic time progresses.
Viewing the expansion of the universe as a process towards holographic equipartition,
he argued that the driving force can be attributed to the difference between the
number of degrees of freedom on a holographic surface and the one in the bulk.
For higher dimensional generalizations and the cases beyond general relativity,
see the further investigations in Refs. \cite{{Cai:2012ip},{Yang:2012wn},{Tu:2013gna},{Ling:2013qoa},{Sheykhi:2013xga},{Sheykhi:2013ffa},{Sheykhi:2013tqa}}.

In this paper, based on the modified proposals in Refs. \cite{Sheykhi:2013xga} and \cite{Yang:2012wn},
we use two methods to derive the modified Friedmann equations when logarithmic and power-law entropic corrections are present.
The results are shown to be consistent with other approaches.

{\emph{Basics.}}---
Consider the metric of a four-dimensional (4D) Friedmann-Robertson-Walker (FRW) universe as follows
\be
ds^2={h}_{ab}dx^{a} dx^{b}+\tilde{r}^2(d\theta^2+\sin^2\theta
d\phi^2),
\ee
Here the 2D metric is  $h_{ab}$ = diag$(-1, \f {a^2} {1-kr^2})$,
and $\tilde{r}=a(t) r$, $x^0=t$, $x^1=r$.
As usual, 
the spatial curvature constant $k = +1$, $0$, and $-1$ correspond to a closed, flat, and open universe,
respectively.

The intriguing discovery of Padmanabhan was based on a hypothetical expansion law \cite{Padmanabhan:2012ik}
\begin{equation}
\frac{d V}{dt}=L_{p}^{2} (N_{\mathrm{sur}}-N_{\mathrm{bulk}}),  \label{dV 0}
\end{equation}
where $V$ is the cosmic volume, $L_{p}$ is the Planck length, and $\triangle N = N_{\mathrm{sur}}-N_{\mathrm{bulk}}$ is the difference between the number of degrees of freedom on a holographic surface and the one in the bulk.
From this, one can derive the Friedmann equations through some basic assumptions and manipulations.
As demonstrated by Cai \cite{Cai:2012ip},
the results for Gauss-Bonnet gravity and Lovelock gravity also agree with
the approach using the first law of thermodynamics (see Ref. \cite{Cai:2005ra}).

Closely following the further discussions of Ref. \cite{Cai:2012ip},
an improved expansion law has been found in Ref. \cite{Sheykhi:2013xga} which is
\begin{equation}
\frac{d \wt V}{dt}=L_{p}^{2} H \wt{r}_A (\wt N_{\mathrm{sur}} - N_{\mathrm{bulk}}).  \label{dV 1}
\end{equation}
Here $\wt V$ is the effective cosmic volume, and $\wt N_{\mathrm{sur}}$ is the effective number of degrees of freedom on the holographic surface.
Furthermore, the apparent horizon radius for the FRW universe is
\begin{equation} \la{ah radius}
 \wt {r}_A=\frac{1}{\sqrt{H^2 + \f {k}{a^2}}},
\end{equation}
where $H = \f {\dot a}{a}$ is the Hubble parameter.
The above new dynamical equation (\ref{dV 1}) allows us to 
deal with the cases where the curvature factor $k$ is not zero.

{\emph{Logarithmic corrections.}}---
We start with the following corrected entropy-area relation
(see e.g. Refs. \cite{{Banerjee:2008cf},{Modesto:2010rm}})
\begin{equation}
S= \frac{A}{4G} + \alpha \ln \frac{A}{4G}+\beta \frac{4G}{A},
\end{equation}
where the Newton's constant is $G = L^2_p$, and the area is $A = 4 \pi {\wt r_A}^2$.
Note that the convention here is $k_B = c = \hbar = 1$.
To be more general, we have also included a reciprocal correction term.

The relevant effective area of the holographic surface is defined as
\be
\wt A = A + 4 \a L^2_p \ln \f {A}{4 L^2_p} + \f {16 \b L^4_p}{A}.
\ee
Then the increase in the effective volume can be calculated to be
\bea   \la{la dv}
\f {d \wt V}{d t} &=& \f {\wt r_A}{2} \f {d \wt A}{d t}  \nn  \\
                  &=& 4 \pi {\wt r_A}^2 \dot{\wt r}_A \Big( 1 + \f {\a L^2_p}{\pi {\wt r_A}^2} - \f {\b L^4_p}{\pi^2 {\wt r_A}^4} \Big).
\eea
By noting the fact
\be
\f {d \wt V}{d t} = - 2 \pi {\wt r_A}^5 \f {d}{d t} \Big( \f 1 {{\wt r_A}^2} +
\f {\a L^2_p}{2 \pi {\wt r_A}^4} - \f {\b L^4_p}{3 \pi^2 {\wt r_A}^6} \Big),
\ee
we propose that the effective number of degrees of freedom on the apparent horizon is
\be    \la{la ns}
\wt N_{\mathrm{sur}} = \f {4 \pi {\wt r_A}^2}{L^2_p} \Big( 1 + \f {\a L^2_p}{2 \pi {\wt r_A}^2}
- \f {\b L^4_p}{3 \pi^2 {\wt r_A}^4} \Big).
\ee

On the other hand, the number of degrees of freedom in the bulk is the same as previous studies.
According to the equipartition law, we have
\bea
N_{\mathrm{bulk}} &=& \frac{2}{T} |E_{\mathrm{Komar}}| = - 2 (\rho +3p) \f {V}{T},
\eea
where the expression of the Komar energy has been inserted.
The appearance of a minus sign is due to the fact that we are considering the accelerating phase here.

Noting that the Hawking temperature \cite{Cai:2008gw} is $T = \f 1{2 \pi \wt r_A}$,
and the cosmic volume is $V = \f {4 \pi {\wt r_A}^3}{3}$, we obtain
\bea  \la{la nb}
N_{\mathrm{bulk}} &=& - \f {16 \pi^2}{3} (\rho +3p) {\wt r_A}^4  \nn \\
                  &=& \f {16 \pi^2}{3} \Big(\f {\dot \r}{H} + 2 \r \Big) {\wt r_A}^4.
\eea
Here the continuity equation $\dot \r + 3 H (\r + p) = 0$ has been applied.

Substituting Eqs. (\ref{la dv}), (\ref{la ns}), and (\ref{la nb}) into the expansion law (\ref{dV 1}), we arrive at
\bea
4 \pi {\wt r_A}^3 \f {\dot a}{a} \Big( 1 + \f {\a L^2_p}{2 \pi {\wt r_A}^2} &-& \f {\b L^4_p}{3 \pi^2 {\wt r_A}^4} \Big)
- 4 \pi {\wt r_A}^2 \dot{\wt r}_A \Big( 1 + \f {\a L^2_p}{\pi {\wt r_A}^2} - \f {\b L^4_p}{\pi^2 {\wt r_A}^4} \Big)  \nn \\
&=& \f {16 \pi^2 L^2_p}{3} \f 1{a} (\dot \r a + 2 \r \dot a) {\wt r_A}^5.
\eea
After multiplying both sides by $\f {a^2}{2 \pi {\wt r_A}^5}$,
it can be rewritten as
\be
\f {d}{dt} \bigg[ a^2 \Big( \f 1 {{\wt r_A}^2} +
\f {\a L^2_p}{2 \pi {\wt r_A}^4} - \f {\b L^4_p}{3 \pi^2 {\wt r_A}^6} \Big) \bigg]
 = \f {8 \pi L^2_p}{3} \f {d}{dt} (\r a^2).
\ee

Integrating the above equation and setting the integration constant to zero,
we find the following modified Friedmann equation
\be
H^2 +\f{k}{a^2} + \f{\a L^2_p}{2\pi}   \Big(H^2+\frac{k}{a^2}\Big)^2
 - \f{\b L^4_p}{3\pi^2} \Big(H^2+\f{k}{a^2}\Big)^3
 = \f{8\pi L^2_p}{3}\r.
\ee
Here the expression of the apparent horizon radius (\ref{ah radius}) has been used.
Thus we have reproduced the result in Ref. \cite{Cai:2008ys}.

{\emph{Power-law corrections.}}---
By studying the entanglement of quantum fields between inside and outside of the horizon,
Ref. \cite{Das:2007mj} found the entropy with power-law entropic corrections as follows
\be
S = \f {A}{4 L^2_p} \big( 1 - K_\a A^{1-\f \a{2}} \big).
\ee
Here $\a$ is a dimensionless parameter (not to be confused with the $\a$ in the above discussions), and
\be
K_\a = \f {\a (4 \pi)^{\f \a{2}-1}} {(4-\a) r^{2-\a}_c},
\ee
where $r_c$ is the crossover scale.

In this case, we can define the effective area of the holographic surface as
\be
\wt A = A - K_\a A^{2 - \f {\a}{2}}.
\ee
The increase in the effective volume is easily found to be
\bea     \la{pl dv}
\f {d \wt V}{d t} &=& \f {\wt r_A}{2} \f {d \wt A}{d t}  \nn  \\
                  &=& 4 \pi {\wt r_A}^2 \dot{\wt r}_A \bigg[ 1 - \f {\a}{2}
                   \Big( \f {r_c}{\wt r_A} \Big)^{\a-2} \bigg].
\eea
Because it can be rewritten as
\be    \la{pl C}
\f {d \wt V}{d t} = - 2 \pi {\wt r_A}^5 \f {d}{d t} \Big( \f 1{{\wt r_A}^2} - \f {r^{\a-2}_c}{{\wt r_A}^\a} + C \Big),
\ee
we assume that the effective number of degrees of freedom on the apparent horizon takes the following form
\be     \la{pl ns}
\wt N_{\mathrm{sur}} = \f {4 \pi {\wt r_A}^2}{L^2_p} \bigg[ 1 - \Big( \f {r_c}{\wt r_A} \Big)^{\a-2} + C {\wt r_A}^2 \bigg],
\ee
where $C$ is a constant to be determined later.

By inserting Eqs. (\ref{pl dv}), (\ref{pl ns}), and (\ref{la nb}) into the expansion law (\ref{dV 1}), we have
\bea
4 \pi {\wt r_A}^3 \f {\dot a}{a} \bigg[ 1 - \Big( \f {r_c}{\wt r_A} \Big)^{\a-2} &+& C {\wt r_A}^2 \bigg]
- 4 \pi {\wt r_A}^2 \dot{\wt r}_A \bigg[ 1 - \f {\a}{2} \Big( \f {r_c}{\wt r_A} \Big)^{\a-2} \bigg]  \nn \\
&=& \f {16 \pi^2 L^2_p}{3} \f 1{a} (\dot \r a + 2 \r \dot a) {\wt r_A}^5.
\eea
Multiplying both sides by $\f {a^2}{2 \pi {\wt r_A}^5}$, we obtain
\be     \la{pl ife}
\f {d}{dt} \bigg[ a^2 \Big( \f 1 {{\wt r_A}^2} - \f {r^{\a-2}_c}{{\wt r_A}^\a} + C \Big) \bigg]
 = \f {8 \pi L^2_p}{3} \f {d}{dt} (\r a^2).
\ee

In the $\a \ra 0$ limit, it should reduce to the usual expression with no entropic corrections
\be
\f {d}{dt} \bigg( \f {a^2}{{\wt r_A}^2} \bigg)
 = \f {8 \pi L^2_p}{3} \f {d}{dt} (\r a^2).
\ee
So the constant is $C = \f 1{r^2_c}$.
Integrating Eq. (\ref{pl ife}) and recalling the expression of the apparent horizon radius in Eq. (\ref{ah radius}),
we arrive at
\be
H^2 +\f{k}{a^2} - \f 1{r^2_c} \bigg[ r^\a_c \Big( H^2 +\f{k}{a^2} \Big)^{\f \a{2}} - 1 \bigg]
 = \f{8\pi L^2_p}{3}\r.
\ee
This is exactly the modified Friedmann equation derived in Refs. \cite{{Sheykhi:2010yq},{Karami:2010bg}}.

{\emph{Generalized expansion law.}}---
Instead of introducing the notion of effective area, one still has another way to obtain the same results.
As shown in Ref. \cite{Yang:2012wn}, we can generalize the expansion law (\ref{dV 0}) as follows
\begin{equation}
\frac{d V}{dt}=L_{p}^{2} f(\triangle N, N_{\mathrm{sur}}).  \label{dV 2}
\end{equation}
This proposal guarantees that the number of degrees of freedom on a holographic surface
is strictly proportional to its area.
On the other hand, the emergence of cosmic space would also depend on the properties of the surface.

To incorporate the cases with nonzero spatial curvature,
we introduce a modified version of the generalized expansion law (\ref{dV 2}) as
\begin{equation}
\frac{d V}{dt}=L_{p}^{2} H \wt{r}_A f(\triangle N, N_{\mathrm{sur}}).  \label{dV 2y}
\end{equation}

Any corrected entropy takes the following form
\be
\wt S = S + s(A).
\ee
We further define
\bea
\wt s(A) &=& 4 L^2_p s(A),   \\
B &=& \f {d}{d A} \wt s(A),   \\
D &=& - 2 {\wt r_A}^2 \int \f {d {\wt r}_A} {{\wt r_A}^3} B.
\eea
Note that in special cases,
the expression of $D$ may have some ambiguity related to the integration constant as in Eq. (\ref{pl C}).

By comparing Eqs. (\ref{dV 1}) and (\ref{dV 2y}),
we can determine the function in the new expansion law as follows
\be   \la{gel f}
f(\triangle N, N_{\mathrm{sur}}) = \f {\triangle N + D N_{\mathrm{sur}}}{1 + B}.
\ee
Using the previous calculations, the parameters for logarithmic and reciprocal corrections can be found to be
\bea
B_L &=& \f {\a L^2_p}{\pi {\wt r_A}^2} - \f {\b L^4_p}{\pi^2 {\wt r_A}^4},    \\
D_L &=& \f {\a L^2_p}{2 \pi {\wt r_A}^2} - \f {\b L^4_p}{3 \pi^2 {\wt r_A}^4}.
\eea
For the power-law corrections, we have
\bea
B_P &=& - \f {\a}{2} \Big( \f {r_c}{\wt r_A} \Big)^{\a-2},   \\
D_P &=& - \Big( \f {r_c}{\wt r_A} \Big)^{\a-2} + \f {{\wt r_A}^2} {r^2_c}.
\eea

One can check that for the 4D Gauss-Bonnet gravity and Lovelock gravity with $k=0$,
Eq. (\ref{gel f}) leads to the same functions as in Ref. \cite{Yang:2012wn}.
This general formula also applies to the various cases studied in Refs. \cite{{Tu:2013gna},{Ling:2013qoa}}.

{\emph{Conclusion.}}---
To sum up,
we applied a modified proposal \cite{Sheykhi:2013xga} of Padmanabhan's emergent cosmic space scenario \cite{Padmanabhan:2012ik}
to the cases where logarithmic and power-law entropic corrections are present,
and deduced the corresponding modified Friedmann equations.
Furthermore,
we gave a modified version of the generalized expansion law found in Ref. \cite{Yang:2012wn}.
This constituted another method to find the Friedmann equations
without the need to introduce the notion of effective area as in Ref. \cite{Cai:2012ip}.
Besides, the resulting general formula (\ref{gel f}) may be beneficial for the investigations of other situations.

As for further directions, it would be interesting to find more connections between
the modified proposals in Refs. \cite{Yang:2012wn} (see also Eq. (\ref{dV 2y})) and \cite{Sheykhi:2013xga}.
Secondly,
a clearer explanation for the procedure used in Refs. \cite{{Sheykhi:2013xga},{Cai:2012ip}}
to obtain the effective number of degrees of freedom on the holographic surface
is desired.
Finally,
in some particular cases, different entropic correction terms which have different origins
lead to similar modified Friedmann equations.
This phenomenon may have deeper physical significance.

{\emph{Acknowledgments}.---
F.-F.Y. would like to thank Peng Huang for long-term discussions about dark energy.
We also thank Rong-Gen Cai and Leonardo Modesto for correspondence.
We later learned of the work in Ref. \cite{Sheykhi:2013ffa} which had some overlap with our results.
This work was supported by National Natural Science
Foundation of China (No. 11275017 and No. 11173028).

\end{document}